\documentclass[12pt]{article}
\usepackage{graphicx}
\usepackage{amsmath}
\newcommand{\Hd}{H^\dagger}

\begin{document}

\baselineskip 0.7cm

\begin{titlepage}
\begin{flushright}
UT-07-06
\end{flushright}
\vskip 1.35cm
\begin{center}
{\large \bf Solutions to large $B$ and $L$ breaking in the Randall-Sundrum model}
\vskip 1.2cm
H. Nakajima and Y. Shinbara
\vskip 0.4cm
\textit{Department of Physics, University of Tokyo, \\ Tokyo 113-0033, Japan}
\vskip 1.5cm
\abstract{
  The stability of proton and neutrino masses
  are discussed in the Randall-Sundrum model.
  We show that relevant operators should be suppressed,
  if the hierarchical Yukawa matrices are explained
  only by configurations of wavefunctions
  for fermions and the Higgs field along the extra dimension.
  We assume a $Z_N$ discrete gauge symmetry to suppress those operators.
  In the Dirac neutrino case,
  there is an infinite number of symmetries
  which may forbid the dangerous operators.
  In the Majorana neutrino case,
  the discrete gauge symmetries should originate from $U(1)_X$ gauge symmetries
  which are broken on the Planck brane.
  We also comment on the $n-\bar{n}$ oscillation as a phenomenon
  which can distinguish those discrete gauge symmetries.
}
\end{center}
\end{titlepage}

\setcounter{page}{2}

\section{Introduction}
Models with low scale quantum gravity \cite{Arkani-Hamed:1998rs, Randall:1999ee}
have been intensely studied, since they can account for the hierarchy
between the electroweak and the fundamental scales.
However, in the low cut-off theories,
higher dimensional operators induce in general
fast proton decays or too large neutrino masses,
since these operators are suppressed only by the low cut-off scale $\sim 1$ TeV
\footnote{
  For other constraints
 (from flavor changing neutral currents,
  precision electroweak measurements, etc.),
  see \cite{Moreau:2006np} and references therein.
}.
It has been argued that these problems can be solved
by some symmetries \cite{Arkani-Hamed:1998sj}
or field configurations along the extra dimensions \cite{Arkani-Hamed:1999dc}.

The solution by field configurations
along the extra dimensions is interesting,
since it also explains the hierarchical Yukawa matrices
without introducing extra spectator fields
\cite{Arkani-Hamed:1999dc, Dvali:2000ha, Gherghetta:2000qt}.
In fact, there are some attempts in higher dimensional models
to explain the hierarchical Yukawa matrices
\cite{Kaplan:2000av, Huber:2000ie, Chang:2005ya}
and sufficiently long proton lifetime
\cite{Arkani-Hamed:1999dc}
by field configurations along the extra dimensions.
However, these benefits do not go together in general,
since the observed fermion masses require sufficient overlaps
of wave functions for fermions,
while the long proton lifetime requires small overlaps.
For example, the proton lifetime is too short in the Randall-Sundrum (RS) model,
if the observed fermion masses are explained by particular field configurations
\cite{Huber:2000ie}.

The purpose of this paper is to study these problems in the RS model.
In the next section, we analyze the proton lifetime and neutrino masses.
We show that relevant operators must be suppressed substantially by hand,
if the hierarchical Yukawa matrices are explained only by field configurations.
In section \ref{se:dis}, we look for symmetries
which may suppress those operators naturally.
Here, we focus on discrete gauge symmetries \cite{Krauss:1988zc}
\footnote{
  Discrete gauge symmeties in extra dimension models
  have been studied in \cite{Davoudiasl:2005ks}.
  However, their model is different from ours in some ways.
  For example, right-handed neutrinos do not acquire
  large Majorana masses in their model.
  In our model, they acquire large Majorana masses
  and the usual seesaw mechanism \cite{seesaw} is realized,
  as we will show in section 3.2.
},
since global symmetries may be explicitly broken
by the topological effects of gravity \cite{Coleman:1988cy},
and there is no continuous gauge symmetry
which supresses the dangerous operators.
We introduce a $Z_N$ discrete gauge symmetry to supress those operators.
In the Dirac neutrino case,
there is an infinite number of symmetries
which may forbid the dangerous operators.
In the Majorana neutrino case,
discrete gauge symmetries cannot forbid the dangerous operators.
However, if the discrete gauge symmetries originate from $U(1)_X$ gauge symmetries,
and they are broken on the Planck brane,
then the dangerous operators can be suppressed.
Finally, we comment on $n-\bar{n}$ oscillation as a phenomenon
which distinguishes between those discrete gauge symmetries.

\section{Higher dimensional operators in the RS model}
\label{se:higher}
In some extra dimension models,
the configurations of fermions and the Higgs field along the extra dimensions
are employed to explain the proton stability,
neutrino masses, and the hierarchical Yukawa matrices.
However, in the RS model,
the field configurations by themselves can not explain
the above issues simultaneously,
as we will show in this section.

First, we summarize our setup.
The metric of the RS model is
\begin{align}
  ds^2 = e^{-2\sigma} \eta_{\mu\nu} dx^\mu dx^\nu + dy^2,
\end{align}
where $\sigma = k|y|$,
and $k \sim M_P = 2.4 \times 10^{18}$ GeV is the AdS curvature.
The fifth dimension $y$ is compactified on an orbifold $S^1/Z_2$.
Two 3-branes reside at the fixed points $y=0$, $y=\pi R$,
which are referred to as the Planck brane and the TeV brane, respectively.
The only scale that appears in this model is the Planck scale:
all terms in the 5D action are characterized by $M_P$.
The effective scale on the TeV brane is $M_T \equiv e^{-\pi kR} M_P$.
We take $M_T \sim 10$ TeV, that is, $kR \sim 10$ in the following arguments.
We put the Higgs field $H$ on the TeV brane to solve the hierarchy problem.
The 5D Dirac fermions
$\Psi_{qi}$, $\Psi_{ui}$, $\Psi_{di}$, $\Psi_{li}$ and $\Psi_{ei}$
are in the bulk,
and their chiral zero modes
$q_{Li} = (u_{Li}, d_{Li})$, $u_{Ri}$, $d_{Ri}$, $l_{Li} = (\nu_{Li}, e_{Li})$ and $e_{Ri}$
are the quarks and leptons of the standard model.
The lower sufficies $i=1,2,3$ denote the generations of fermions.
The configurations of fermions explain the hierarchical Yukawa matrices
\cite{Gherghetta:2000qt}.

The operators which generate 4D Majorana neutrino masses are
\begin{align}
       \int d^4x \int dy \sqrt{-g} \frac{1}{M_P^2}
       H H \bar{\Psi}_{li}^{c} \Psi_{lj} \delta(y-\pi R)
\equiv \int d^4x (m_\nu)_{ij} \bar{\nu}_{Li}^{c} \nu_{Lj}
     + \dots,
\end{align}
where the upper suffix $c$ denotes the charge conjugation.
The 4D Majorana neutrino masses $(m_\nu)_{ij}$ are given by
\begin{align}
  (m_\nu)_{ij} = \frac{v^2}{M_T} T(c_{li}) T(c_{lj}),
\end{align}
where $v \equiv e^{-\pi kR} \langle H \rangle \sim 100$ GeV
is the VEV of the Higgs field, and $T(c)$ is
\begin{align}
     T(c)
\sim \left\{
       \begin{array}{ll}
         \sqrt{1/2-c}                          & \textrm{for $c<1/2$} \\
         \sqrt{c-1/2} \times (M_T/M_P)^{c-1/2} & \textrm{for $c>1/2$}
       \end{array}
     \right.
\end{align}
(see Appendix \ref{app} for details).

The 5D Yukawa interactions which generate 4D Dirac mass terms are
\begin{align}
       \int d^4x \int dy \sqrt{-g} \frac{(y_e)_{ij}}{M_P}
       H \bar{\Psi}_{li} \Psi_{ej} \delta(y-\pi R)
\equiv \int d^4x (m_e)_{ij} \bar{e}_{Li} e_{Rj}
     + \dots,
\end{align}
where $(y_e)_{ij} \sim O(1)$ are dimensionless 5D Yukawa couplings.
The 4D Dirac mass matrix $(m_e)_{ij}$ is given by
\begin{align}
  (m_e)_{ij} = v \times T(c_{li}) (y_e)_{ij} T(c_{ej}).
\end{align}
Using $m_\tau \sim (m_e)_{33} \sim 1$ GeV, we have
\begin{align}
       m_{\nu \tau}
  \sim (m_\nu)_{33}
& =    \frac{v^2}{M_T} T(c_{l3}) T(c_{l3}) \notag \\
& \sim \frac{v^2}{M_T} \left\{ \frac{m_\tau}{v (y_e)_{33} \, T(c_{e3})} \right\}^2
> 100 \textrm{ keV},
\end{align}
where we have used $(y_e)_{33} \sim 1$ and $T(c) < 1$ for $c \sim 1$.
This neutrino mass is well above the limit $\sum m_\nu < 0.68$ eV
obtained from the WMAP observations \cite{Spergel:2006hy}
and the implications of atmospheric neutrino oscillation
$m_{\nu_3}^2 \sim \Delta m_\mathrm{atm}^2 \sim 10^{-3} \textrm{ eV}^2$
\cite{Ashie:2005ik}.

Dimension 6 operators
\begin{align}
  \hat{u}_{L1}\hat{d}_{L1}\hat{u}_{R1}\hat{e}_{R1}, \quad
  \hat{u}_{R1}\hat{d}_{R1}\hat{u}_{L1}\hat{e}_{L1}, \quad \cdots,
\label{proton}
\end{align}
lead to an unacceptably short proton lifetime,
if they are not suppressed sufficiently.
Here $\hat{u}_{L,R}$, $\hat{d}_{L,R}$ and $\hat{e}_{L,R}$
denote mass eigenstates of fermions.
They are related to the electroweak eigenstates
$u_{L,R}$, $d_{L,R}$ and $e_{L,R}$ via
\begin{align}
  & u_{Li} = \sum_j U^L_{ij} \hat{u}_{Lj}, &
  & d_{Li} = \sum_j D^L_{ij} \hat{d}_{Lj}, &
  & e_{Li} = \sum_j E^L_{ij} \hat{e}_{Lj}, & \notag \\
  & u_{Ri} = \sum_j U^R_{ij} \hat{u}_{Rj}, &
  & d_{Ri} = \sum_j D^R_{ij} \hat{d}_{Rj}, &
  & e_{Ri} = \sum_j E^R_{ij} \hat{e}_{Rj}, &
  \label{01}
\end{align}
where $U^L$, $U^R$, $D^L$, $D^R$, $E^L$ and $E^R$
are $3 \times 3$ unitary matrices.
The eigenvalues of fermion mass matrices are
\newcommand{\Ld}{{L\dagger}}
\newcommand{\Rd}{{R\dagger}}
\begin{align}
    (m_u)_i
& = v \times \sum_{j,k} (U^\Rd)_{ij} T(c_{uj}) (y_u)_{jk} T(c_{qk}) U^L_{ki}, \notag \\
    (m_d)_i
& = v \times \sum_{j,k} (D^\Rd)_{ij} T(c_{dj}) (y_d)_{jk} T(c_{qk}) D^L_{ki}, \notag \\
    (m_e)_i
& = v \times \sum_{j,k} (E^\Rd)_{ij} T(c_{ej}) (y_e)_{jk} T(c_{lk}) E^L_{ki},
    \label{04}
\end{align}
where dimensionless 5D Yukawa couplings $y_u$, $y_d$ and $y_e$ are $\sim O(1)$.

The operator $\hat{u}_{L1} \hat{d}_{L1} \hat{u}_{R1} \hat{e}_{R1}$
is a mixing of operators $u_{Lh} d_{Li} u_{Rj} e_{Rk}$.
The effective suppression scales of the operators $u_{Lh} d_{Li} u_{Rj} e_{Rk}$ are
\begin{align}
    \frac{1}{M(u_{Lh} d_{Li} u_{Rj} e_{Rk})^2}
& = \frac{1}{M_T^2} \ T(c_{qh}) T(c_{qi}) T(c_{uj}) T(c_{ek}) \notag \\
& - \frac{1}{M_P^2} \ P(c_{qh}) P(c_{qi}) P(c_{uj}) P(c_{ek}),
    \label{03}
\end{align}
where
\begin{align}
     P(c)
\sim \left\{
       \begin{array}{ll}
         \sqrt{1/2-c} \times (M_T/M_P)^{1/2-c} & \textrm{for $c<1/2$} \\
         \sqrt{c-1/2}                          & \textrm{for $c>1/2$}
       \end{array}
     \right.
     \label{P(c)}
\end{align}
(see Appendix \ref{app} for details).
Taking the mixing into account,
the suppression scales of the operators
$\hat{u}_{L1} \hat{d}_{L1} \hat{u}_{R1} \hat{e}_{R1}$ and
$\hat{u}_{R1} \hat{d}_{R1} \hat{u}_{L1} \hat{e}_{L1}$ are
\begin{align}
    \frac{1}{M(\hat{u}_{L1} \hat{d}_{L1} \hat{u}_{R1} \hat{e}_{R1})^2}
& = \sum_{h,i,j,k}
    \frac{U^L_{h1} D^L_{i1} U^R_{j1} E^R_{k1}}{M(u_{Lh} d_{Li} u_{Rj} e_{Rk})^2},
    \notag \\
    \frac{1}{M(\hat{u}_{R1} \hat{d}_{R1} \hat{u}_{L1} \hat{e}_{L1})^2}
& = \sum_{h,i,j,k}
    \frac{U^R_{h1} D^R_{i1} U^L_{j1} E^L_{k1}}{M(u_{Rh} d_{Ri} u_{Lj} e_{Lk})^2}.
    \label{02}
\end{align}
Using Eq.(\ref{03}), we see that these summations are approximately given by
\footnote{
  There would be sets $\{ i,j,k,l \}$ for which
  the second term in Eq.(\ref{03}) dominates.
  However, contributions from these sets
  to the summation in Eq.(\ref{02}) are negligible.
  If the second term in Eq.(\ref{03}) dominates for all sets $\{ i,j,k,l \}$,
  then the factors $T(c)$ are too small to explain the fermion masses.
}
\begin{align}
    \frac{1}{M(\hat{u}_{L1} \hat{d}_{L1} \hat{u}_{R1} \hat{e}_{R1})^2}
&   \simeq
    \sum_{h,i,j,k}
    \frac{1}{M_T^2}
    \, T(c_{qh}) T(c_{qi}) T(c_{uj}) T(c_{ek})
    U^L_{h1} D^L_{i1} U^R_{j1} E^R_{k1},
    \notag \\
    \frac{1}{M(\hat{u}_{L1} \hat{d}_{L1} \hat{u}_{R1} \hat{e}_{R1})^2}
&   \simeq
    \sum_{h,i,j,k}
    \frac{1}{M_T^2}
    \, T(c_{uh}) T(c_{di}) T(c_{qj}) T(c_{lk})
    U^R_{h1} D^R_{i1} U^L_{j1} E^L_{k1}.
\end{align}
Thus we have
\begin{align}
 & \hspace{-45pt}
   \left[
   \frac{1}{M(\hat{u}_{L1} \hat{d}_{L1} \hat{u}_{R1} \hat{e}_{R1})^2}
   \right]^*
   \frac{1}{M(\hat{u}_{R1} \hat{d}_{R1} \hat{u}_{L1} \hat{e}_{L1})^2}
   \notag \\
   \simeq
   \frac{1}{M_T^2} \frac{1}{M_T^2}
 & \sum_{h,h'} (U^\Ld)_{1h} T(c_{qh}) T(c_{uh'}) U^R_{h'1}
   \sum_{i,i'} (D^\Ld)_{1i} T(c_{qi}) T(c_{di'}) D^R_{i'1}
   \notag \\
 & \sum_{j,j'} (U^\Rd)_{1j} T(c_{uj}) T(c_{qj'}) U^L_{j'1}
   \sum_{k,k'} (E^\Rd)_{1k} T(c_{ek}) T(c_{lk'}) E^L_{k'1}.
\end{align}
Since $y_u$, $y_d$ and $y_e$ are $\sim O(1)$ in Eq.(\ref{04}), we have
\begin{align}
         \left[
         \frac{1}{M(\hat{u}_{L1} \hat{d}_{L1} \hat{u}_{R1} \hat{e}_{R1})^2}
         \right]^*
         \frac{1}{M(\hat{u}_{R1} \hat{d}_{R1} \hat{u}_{L1} \hat{e}_{L1})^2}
\sim \ & \frac{1}{M_T^2} \frac{1}{M_T^2}
         \frac{(m_u)_1^*}{v} \frac{(m_d)_1^*}{v} \frac{(m_u)_1}{v} \frac{(m_e)_1}{v}
         \notag \\
\sim \ & \frac{1}{(\textrm{$10^9$ GeV})^4}.
         \label{05}
\end{align}

The decay rates of $p \to \pi^0 e^+$ induced by the dimension 6 operators
$O_1 = \hat{u}_{R1} \hat{d}_{R1} \hat{u}_{L1} \hat{e}_{L1}$ and
$O_2 = \hat{u}_{L1} \hat{d}_{L1} \hat{u}_{R1} \hat{e}_{R1}$ are
\begin{align}
  \Gamma (p \to \pi^0 e^+)
= \sum_{i=1,2}
  \frac{1}{M(O_i)^4} \frac{1}{4 \pi^2}
  \int \frac{d^3 q'}{2q_0} \int \frac{d^3 k'}{2k_0}
  |W(k-q')|^2 \frac{k'k}{2m_p} \delta^4 (k-k'-q'),
  \label{decay_rate}
\end{align}
where $W(k-q')$ is the form factor of the $p \to \pi$ matrix element
\begin{align}
  \langle \pi(q) | (\hat{d}_{R1} \hat{u}_{R1}) \hat{u}_{L1} | p(k) \rangle
= W(q-k) p_L(k),
\end{align}
and $k$, $k'$ and $q$ are four momenta of proton, positron and pion, respectively.
The momentum dependence of $W$ is weak and $W \simeq -0.15 \textrm{ GeV}^2$
\cite{Aoki:1999tw}, so that
\begin{align}
       \tau(p \to \pi^0 e^+)
\simeq 7.5
\times 10^{31}
\times \left( \frac{\min[M(O_i)]}{\textrm{$10^{15}$ GeV}} \right)^4
\times \left( \frac{\textrm{0.15 GeV$^2$}}{|W|} \right)^2
       \textrm{yr.}
\end{align}
Thus the suppression scales obtained in Eq.(\ref{05}) are too small
to explain the observed proton lifetime.

We have seen that operators concerning
with the Majorana neutrino masses and the proton decay
should be suppressed by small factors, or forbidden by some symmetries.
We consider that the former solution contradicts with the philosophy of the RS model,
that is, solving the hierarchy problem without fine tunings.
In the next section, we look for the symmetries which may suppress these operators.

\section{Discrete gauge symmetry}\label{se:dis}
In this section, we look for the symmetries
which may suppress the dangerous operators.
We concentrate on gauge symmetries,
since any global symmetries may be explicitly broken
by the topological effects of gravity \cite{Coleman:1988cy}.
In addition, we consider discrete gauge symmetries,
since there is no continuous anomaly-free symmetry except for $U(1)_{B-L}$,
which can not supress the dangerous operators for the proton decay.
We introduce only one discrete gauge symmetry:
the gauge symmetry of the action is assumed to be
$SU(3)_C \times SU(2)_L \times U(1)_Y \times Z_N$,
where $SU(3)_C \times SU(2)_L \times U(1)_Y$
is the gauge group of the standard model.
Furthermore, we introduce 5D Dirac fields $\Psi_{\nu i}$ $(i=1,2,3)$
to obtain acceptable neutrino masses.
The zero modes of $\Psi_{\nu i}$ are the right-handed neutrinos $\nu_{Ri}$.

\subsection{$Z_N$ symmetry for the Dirac neutrino case}\label{Dirac}
Here we discuss $Z_N$ discrete gauge symmetries in the Dirac neutrino case.
Those $Z_N$ symmetries must respect the Yukawa terms.
The $Z_N$ charge of each field is constrained as Table \ref{yukawa}.
We can set the $Z_N$ charge of the Higgs field to $0$
without loss of generality by using a gauge rotation of $U(1)_Y$
in the standard model.

\renewcommand{\arraystretch}{1.2}
\begin{table}[tbp]
\begin{center}
\begin{tabular}{|c|ccccccc|}
  \hline
        & $q_L$ & $u_R$ & $d_R$ & $l_L$ & $e_R$ & $\nu_R$ & $H$ \\
  \hline
  $Z_N$ &  $m$  &  $m$  &  $m$  &  $p$  &  $p$  &   $p$   & $0$ \\
  \hline
\end{tabular}
\caption{
  $Z_N$ charges consistent with the Yukawa interactions.
  Here we set the charge of the Higgs field to $0$
  by using a gauge rotation of $U(1)_Y$.
}
\label{yukawa}
\end{center}
\end{table}

The anomaly cancellation conditions which include $Z_N$ are
\begin{align}
  \left\{
  \begin{array}{rcl}
                              0 & = & \dfrac{   1}{2} r_1 N, \medskip \\
    \dfrac{9}{2}m+\dfrac{3}{2}p & = & \dfrac{   1}{2} r_2 N, \medskip \\
                              0 & = & \dfrac{\eta}{2} r_3 N + r_4 N,
  \end{array}
  \right.
\end{align}
where $r_{i}$ are integers and $\eta=1,0$ for $N=$ even, odd.
The first equation comes from the cancellation of $\{Z_N\}\{SU(3)_C\}^2$ anomalies,
the second from the cancellation of $\{Z_N\}\{SU(2)_L\}^2$ anomalies,
and the last from the cancellation of $Z_N$-gravitational anomalies.
Here we omit the cancellation of
$\{Z_N\}^3$, $\{Z_N\}^2 \{U(1)_Y\}$ and $\{Z_N\} \{U(1)_Y\}^2$ anomalies,
because these constraints are always satisfied
by adding heavy particles with appropriate charges
\cite{Ibanez:1991hv, Banks:1991xj}.
There is an infinite number of $Z_N(p,m)$ discrete gauge symmetries
which satisfy these constraints.

Now we examine which operators should be suppressed by symmetries.
First, we consider operators with dimension $n \ge 12$.
The proton lifetime derived from a dimension $n$ operator is roughly estimated as
\begin{align}
  \tau_p \sim \frac{M_n^{2n-8}}{m_p^{2n-7}},
\end{align}
where $M_n$ is the suppression scale of the proton decay operator.
For $n \ge 12$, $\tau_p$ is longer than the experimental bound
$\tau_p > 1.9 \times 10^{29}$ years for $M_n = M_T = 10$ TeV.
The other $B$, $L$ breaking operators with dimension $n \ge 12$
would also be sufficiently suppressed by $M_T$
\footnote{
  There may be operators with dimension $n \ge 12$
  which require supressions $M_n > 10$ TeV.
  Even in that case, we can forbid them by choosing appropriate symmetries,
  or suppress them by tuning the fermion configurations.
  We can also suppress them by assuming that
  $Z_N$ symmetries originate from $U(1)_X$ symmetries,
  as we will see in the next subsection.
}.

Thus we concentrate on the $B$, $L$ breaking operators with dimension $n<12$,
which include the dimension $6$ proton decay operators
and the dimension 5 neutrino mass opertors discussed in section \ref{se:higher}.
The $Z_n$ charges of those operators
are $2p$, $4p$, $6p$, $3m \pm p$, $3m \pm 3p$ and $6m$,
and an infinite number of symmetries forbid these operators.
For example, an infinite series of $Z_{9m+3}(1,m)$ $(m = 2, 3, \dots)$ symmetries 
\footnote{
  The symmetries in this series are independent of each other
  for the following reason.
  Equivalent discrete gauge symmetries are related
  through the charge conjugation or $U(1)_Y$ gauge rotation.
  Under the convention of Table \ref{yukawa},
  any symmetry equivalent to $Z_N(p,m)$
  takes the form $Z_{nN}(np,n(m+\frac{kN}{3}))$.
  Thus equivalent symmetries have a common value of $N/p$.
}
are appropriate.
Thus, there is an infinite number of symmetries
which forbid the dangerous operators in the Dirac neutrino case.

\subsection{$Z_N$ symmetry for the Majorana neutrino case}
Let us consider the case where the seesaw mechanism \cite{seesaw}
induces the light Majorana neutrino masses.
Naively, this seems to be impossible,
since the charges of $M \nu_R \nu_R$ are the same
as those of $H H l_L l_L$ which induce too large neutrino masses.
However, this is not the case,
when the discrete gauge symmetries originate from $U(1)_X$ gauge symmetries.

Let us consider that
the scalar field $\Phi$ which breaks the $U(1)_X$ symmetry lives on the Planck brane
\footnote{
  A model in which the lepton number symmetry is broken on the Planck brane
  for the Dirac neutrino case is discussed in \cite{Gherghetta:2003he}.
},
and $\nu_R$ acquire the Majorana masses through the coupling with $\Phi$:
\begin{align}
       \int d^4x \int dy \sqrt{-g}
       \frac{1}{M_P} \Phi \bar{\Psi}_{\nu i}^c \Psi_{\nu j} \delta(y)
\equiv \int d^4x M_{ij} \bar{\nu}_{Ri}^{c} \nu_{Rj}
     + \dots,
\end{align}
where $M_{ij}$ are given by
\begin{align}
  M_{ij} = \langle \Phi \rangle P(c_{\nu i}) P(c_{\nu j}).
\end{align}
Assuming $\langle \Phi \rangle \sim M_P$,
$M_{ij}$ take values between $M_T$ and $ M_P$ for $0 < c_{\nu i}, c_{\nu j} < 1$.

The operators $H H l_L l_L$ also couple with $\Phi$,
since their $U(1)_X$ charges are the same as those of $\nu_R \nu_R$.
Thus the dangerous Majorana neutrino mass terms
appear only through combinations with $\Phi$:
\begin{align}
  \int d^4x \int dy \sqrt{-g} \frac{1}{M_P^4}
  \Phi H H \bar{\Psi}_{li}^c \Psi_{lj} \delta(y) \delta(y-\pi R),
  \label{HHLL}
\end{align}
which should vanish, since $H$ do not overlap with $\Phi$.

Thus, the operators $H H l_L l_L$ induced directly
from the $\Phi$ condensation are negligible.
We now consider Yukawa interaction terms $\Hd \bar{l}_L \nu_R$
and the Majorana mass terms $M \nu_R \nu_R$ to estimate neutrino masses.
A model with these mass terms was suggested in \cite{Huber:2003sf},
and the effective light neutrino masses are approximately given by
\begin{align}
  (m_{\nu})_{ij}
= \frac{\{v \, T(c_{li})T(c_{\nu j})\}^2}
       {\langle \Phi \rangle P(c_{\nu j})P(c_{\nu j})},
\end{align}
which take values between $\sim 1$ GeV and $\sim 10^{-33}$ eV
for $0 < c_{li}, c_{\nu j} < 1$.
Thus observations concerning with neutrino masses are easily explained.

Let us count the $Z_N$ discrete gauge symmetries
which satisfy the anomaly cancellation conditions
and respect the Majorana mass terms
\footnote{
  Discrete R-symmetries which respect the Majorana mass terms
  in MSSM and SUSY GUT were discussed in \cite{Kurosawa:2001iq}.
}.
The anomaly cancellation conditions are the same as those of the last subsection.
Since the Majorana mass terms $M \nu_R \nu_R$ are induced
from the interaction terms of $\Phi$ and $\nu_R$,
there arises another condition for $p$ and $N$,
\begin{align}
  2p = r_5 N,
\end{align}
where $r_5$ is an integer.
Thus the discrete gauge symmetries are
\begin{align}
  Z_{ 1} & (p,m) = (0,0) : \textrm{completely broken } U(1)_X, \notag \\
  Z_{ 2} & (p,m) = (1,1), \notag \\
  Z_{ 9} & (p,m) = (0,2), \notag \\
  Z_{18} & (p,m) = (9,1).
  \label{dis_sym}
\end{align}
There are other discrete symmetries which satisfy these constraints.
However, they are embedded in the above symmetries,
or $U(1)_Y$-gauge equivalents of those \cite{Hamaguchi:1998wm}.

All of these symmetries allow some of the dangerous operators,
whose $Z_N$ charges are $2p$, $4p$, $6p$, $3m \pm p$, $3m \pm 3p$ and $6m$.
However, depending on the $U(1)_X$ charges of fermions,
these operators can be suppressed.
To see this, we analyze general properties
of higher dimensional operators which include $\Phi$.

Consider an operator which consists of $2n$ fermions and a scalar field $\Phi$
\begin{align}
       \int d^4x \int dy \frac{1}{M_{P }^{4n-3}}
       \Phi \Psi_1 \Psi_2 \dots \Psi_{2n} \delta(y)
\equiv \int d^4x         \frac{1}{M_{2n}^{3n-4}}
            \psi_1 \psi_2 \dots \psi_{2n},
\end{align}
where $\Psi_i$ are 5D Dirac fields and $\psi_i$ are their zero modes.
Then the effective suppression scale $M_{2n}$ is given by
\begin{align}
  \frac{1}{M_{2n}^{3n-4}}
= \frac{1}{M_P^{3n-3}} \Phi P(c_1) P(c_2) \dots P(c_{2n})
< \frac{1}{M_P^{3n-4}},
\end{align}
where we used $\langle \Phi \rangle \sim M_P$ and Eq.(\ref{P(c)}).
Thus any higher dimensional operator with nonzero $U(1)_X$ charge
is Planck suppressed in 4D effective theory.

For all discrete symmetries in Eq.(\ref{dis_sym}),
we can set the $U(1)_X$ charges of fermions
so that the dangerous operators have to couple with $\Phi$.
For example, consider the case where the $U(1)_X$ charges
of $\Phi$, $q_L$ and $l_L$ are $1$, $3$ and $2$, respectively.
The $U(1)_X$ symmetry is broken to $Z_{1}(0,0)$ in low energy.
In this case, all the dangerous operators have nonzero $U(1)_X$ charges,
and become Planck suppressed operators.

We comment on $n-\bar{n}$ oscillation as a phenomenon
which would distinguish between the above discrete gauge symmetries.
The $n-\bar{n}$ oscillation is induced
by the dimension 9 operator $(u_R d_R d_R)^2$.
In the case of $Z_2(1,1)$, $Z_9(0,2)$ and $Z_{18}(9,1)$ symmetries,
this operator is always Planck suppressed,
since $u_R$ and $d_R$ have nonzero $U(1)_X$ charges.
In the case of completely broken $U(1)_X$,
the $U(1)_X$ charge of $(u_R d_R d_R)^2$ can be set to zero.
Then the suppression scale is determined by the configurations of quarks,
and could be tuned to the current lower bound,
which is evaluated to be $10^5$ GeV \cite{Mohapatra:1989ze}.
Thus the $n-\bar{n}$ oscillation could be observed in future experiments
\footnote{
  There are proposals to improve the precision constraints
  by at least two orders of magunitude \cite{Kamyshkov:2002vm}.
},
if the discrete gauge symmetry is $Z_1(0,0)$. \bigskip

\section{Conclusion}
It has been argued that short proton lifetime
and too large neutrino masses are most likely induced in the RS model,
if the hierarchical Yukawa matrices are explained
only by the field configurations along the extra dimension.
We have confirmed that the above unwanted phenomenon are inevitable,
and hence searched for the discrete gauge symmetries
which may forbid the dangerous operators
in the Dirac neutrino case and the Majorana neutrino case.
We have found that there is an infinite number
of such symmetries for the Dirac neutrino case.
For the Majorana neutrino case,
the discrete gauge symmetries should originate from $U(1)_X$ gauge symmetries,
and they should be broken on the Planck brane.
Furthermore, if the $U(1)_X$ symmetry is completely broken,
the $n-\bar{n}$ oscillation could be observed in future experiments.

\section*{Acknowledgments}
We thank T. T. Yanagida for stimulating suggestions. 

\appendix
\section{Profile of 5D fermion} \label{app}
Here we summarize the convention for fermions in the bulk \cite{Grossman:1999ra}.
The kinetic and mass terms of a 5D Dirac field $\Psi$ are
\begin{align}
  \int dx^4 \int dy \sqrt{-g}
  \left\{
    \bar{\Psi} i \gamma^A {e_A}^M (\partial_M + \Omega_M) \Psi + m_D \bar{\Psi} \Psi
  \right\},
\end{align}
where ${e_A}^M$ is the vielbein, and $\Omega_M$ is the spin connection.
The 5D Dirac mass $m_D$ is odd, and parametrized as $m_D = c \sigma'$.
The 5D Dirac field $\Psi$ is decomposed to
\begin{align}
  \Psi(x,y)
= \begin{bmatrix}
    \psi_L^{(n)}(x) f_L^{(n)}(y) \\
    \psi_R^{(n)}(x) f_R^{(n)}(y)
  \end{bmatrix},
\end{align}
where $\psi_L^{(n)}$ and $\psi_R^{(n)}$ are 4D Weyl spinors.
A pair of Weyl spinors $(\psi_L^{(n)},\psi_R^{(n)})$
forms a 4D Dirac spinor $\psi^{(n)}$ with mass $m_n \sim M_T$.
The zero mode is chiral and does not have Dirac mass term,
since the $y$ direction is compactified on $S^1/Z_2$.
The chirality of zero mode depends on the $Z_2$ transformation property of $\Psi$.
Here we take the zero mode to be left-handed.
The wavefunction $f_L^{(0)}(y)$ for the chiral zero mode $\psi_L^{(0)}$ is
\begin{align}
  f_L^{(0)}(y) = N_0 e^{(2-c)k|y|}.
\end{align}
The normalization condition is
\begin{align}
  \int dy \, e^{-3k|y|} f_L^{(0)}(y) f_L^{(0)}(y) = 1
\end{align}
and we have
\begin{align}
  N_0 = \sqrt{\frac{(1-2c)k}{2\{e^{(1-2c)k\pi R}-1\}}}.
\end{align}
$T(c)$ and $P(c)$ which we have used in this paper are defined as
\begin{align}
  T(c) & \equiv \frac{1}{\sqrt{k}} \ e^{-(3/2)k|y|} f_L^{(0)}(y)|_{y = \pi R}, \\
  P(c) & \equiv \frac{1}{\sqrt{k}} \ e^{-(3/2)k|y|} f_L^{(0)}(y)|_{y = 0}    .
\end{align}

\end{document}